\documentstyle{aipproc}
\input{epsf}

\begin{document}

\begin{flushright}
\begin{tabular}{l}
FERMILAB--CONF--97/278--T\\
hep-ph/9709448
\end{tabular}
\end{flushright}

\title{\bf Beautiful CP Violation}
\author{Isard Dunietz}
\address{\it Fermi National Accelerator Laboratory, P.O. Box 500,  
Batavia, IL
60510}

\maketitle 
\begin{abstract}
CP violation is observed to date only in $K^0$ decays and is
parameterizable by a single quantity $\epsilon$.  Because it is one of the
least understood phenomena in the Standard Model and holds a clue to
baryogenesis, it must be investigated further. Highly specialized searches in $K^0$
decays are possible. Effects in $B$ decays are much larger.  In addition
to the traditional $B_d \to J/\psi K_S, \pi^+\pi^-$ asymmetries, CP
violation could be searched for in already existing inclusive $B$ data
samples. The rapid $B_s - \overline B_s$ oscillations cancel in untagged
$B_s$ data samples, which therefore allow feasibility studies for the
observation of CP violation and the extraction of CKM elements with
present vertex detectors. The favored method for the extraction of the CKM
angle $\gamma$ is shown to be unfeasible and a solution is presented
involving striking direct CP violation in charged $B$ decays. Novel methods for determining the $B_s$ mixing parameter $\Delta m$ are described without the traditional requirement of flavor-specific final states.
\end{abstract}

\section{Introduction}

More than thirty years after its discovery, CP violation remains a mystery.  Our entire knowledge about it can be summarized by the single parameter $\epsilon$~\cite{winsteinw}.  CP violation is not just a quaint tiny effect observed in $K^0$ decays, but is one of the necessary ingredients for baryogenesis~\cite{sakharov}. Within the CKM model, it is connected also to the quark-mixing and hierarchy of quark masses.  A successful theory of CP violation will have far-reaching ramifications in cosmology and high energy physics.

At present, we are not able to answer even the question raised by Wolfenstein more than 30 years ago:  Is CP violation due to a new superweak interaction, which would show up essentially only in mixing-induced phenomena?  Or are there direct CP violating effects?  There exists a multitude of scenarios for CP violation, all consistent with $\epsilon$.  What is needed is the observation of many independent CP violating effects.  This would be invaluable in directing us toward a more fundamental understanding of CP violation, in analogy to the history of parity violation.  There a variety of measurements guided us to the successful $V-A$ theory~\cite{vminusa}.

Searches for (direct) CP violation in $K$ and hyperon decays are important~\cite{winsteinw,buchallak}.  Because the expected effects are either tiny for processes with sizable BR's or could be large but then involve tiny BR's ${\cal O} (10^{-11})$, ingenious experimental techniques are being developed to overcome those handicaps.

A whole class of additional independent CP measurements can be obtained from studies of $b$-hadron decays.  Although CP violation may not be (entirely) due to the CKM model, that model serves here as a guide. Decays of $b$-hadrons can access large CKM phases and thus large CP
violation, because the $b$-quark is a member of the third generation. There are many proposed methods that involve large CP violating effects~\cite{burasf}.  This talk focuses on recently discussed phenomena, some of which can be studied with presently existing data samples.

First, (semi-)inclusive $B$ decays are expected to exhibit CP violation and CKM parameters can be extracted~\cite{sachs,bbdcp,stodolsky}.  Even the $B_s$ mixing-parameter $\Delta m$ could be determined from such flavor-nonspecific final states, in addition to the conventional methods~\cite{snowmass,xlep}.  Second, untagged $B_s$ data samples are predicted to exhibit CP violation and permit the extraction of CKM parameters, as long as the $B_s$ width difference is significant~\cite{untagged}.  The far-reaching physics potential of the $B_s \to J/\psi \phi$ process is touched upon.  The third topic explains why the favorite method for determining the CKM angle $\gamma$, pioneered by Gronau-London-Wyler (GLW)~\cite{glw}, is unfeasible.  The CKM parameter can be cleanly extracted~\cite{ads}, however, when one incorporates the striking, direct CP violating effects in
$B \to D^0/\overline D^0$ transitions~\cite{stone}, which were not considered by GLW.

\section{Exclusive and Inclusive $B$ Decays}
 
Traditional methods involve exclusive modes such as 
$J/\psi K_S$ \cite{bigisanda}, 
$\pi^+\pi^-$ \cite{wolfenstein,dudw,dunietzr}, and
study the rate-asymmetry between 
\begin{equation} 
B_d (t)\rightarrow
J/\psi K_S,\; \pi^+\pi^- \neq \overline B_d (t)\rightarrow J/\psi K_S,\;
\pi^+\pi^- \;.  
\end{equation} 
The effective BR is tiny $\sim 10^{-5}$,
but the asymmetries are large ${\cal O} (1)$.  How does this large
asymmetry come about? The unmixed $B_d$ could decay into $J/\psi K_S$
directly, $B_d \rightarrow J/\psi K_S$. The CP conjugated process is the
direct decay, $\overline B_d \rightarrow J/\psi K_S$. To excellent
accuracy, those two direct decay rates are equal.  The $B_d$ could mix
first into a $\overline B_d$ and then decay to $J/\psi K_S , B_d (t) 
\rightarrow \overline B_d\rightarrow J/\psi K_S$. The CP conjugated
process is the mixing-induced $\overline B_d (t)\rightarrow B_d\rightarrow
J/\psi K_S$ transition. Again, to excellent accuracy, the magnitudes of
the two mixing-induced amplitudes are the same.  The large CP violation
predicted in the CKM model occurs because of the interference of the
direct and mixing-induced amplitudes. To form the asymmetry, it is not
sufficient to reconstruct the final state $J/\psi K_S$.  One must be able
to distinguish those reconstructed events as originating from an initial
$B_d$ versus $\overline B_d$ (referred to as tagging). 

Initially (at $t=0$) the neutral $B$ meson has no time to mix.  At $t=0$
there is no mixing-induced amplitude and thus no CP violation. There is
almost no loss in measuring the asymmetry by not considering $J/\psi K_S$
events within the first $B_d$ lifetime or so. While the rate is largest during that time-interval,
the asymmetry is tiny and needs large proper times to build itself
up~\cite{dunietzr,dunietzn}. Triggering on detached vertices is thus more
efficient for such CP violation studies than one might think naively. 

Inclusive $B $ samples are many orders of magnitude larger than the
exclusive ones and can be accessed by vertexing.  The
time-dependent, totally inclusive asymmetry,
\begin{equation}
\label{incldef} 
I(t) \equiv \frac{\Gamma (B^0 (t)\rightarrow {\text all})
- \Gamma (\overline B^0 (t)\rightarrow {\text all})}{\Gamma (B^0
(t)\rightarrow {\text all}) + \Gamma (\overline B^0 (t)\rightarrow {\text
all})}\;, 
\end{equation} 
is CP violating~\cite{bbdcp,stodolsky}. That
appears to be rather puzzling, especially because the CPT theorem
guarantees that the totally inclusive width is the same for particle and
antiparticle. That CPT stranglehold is removed, because $B^0 -\overline
B^0$ mixing provides an additional amplitude and thus novel interference
effects.  The totally inclusive CP asymmetry $I(t)$ is related to the
wrong-sign asymmetry~\cite{paistreiman,hitoshi} 
\begin{equation}
\label{wrongsign}
\frac{\Gamma (B^0 (t)\rightarrow W) - \Gamma (\overline B^0
(t)\rightarrow
\overline W)}{\Gamma (B^0 (t)\rightarrow W) + \Gamma (\overline B^0
(t)\rightarrow \overline W)} = - a = - Im \frac{\Gamma_{12}}{M_{12}}\;,
\end{equation} 
where $W$ denotes ``wrong-sign" flavor-specific
modes that come only from $\overline B^0\rightarrow W$ and never from $B^0
\rightarrow W,$ such as $W=\ell^- X$ and $W=D_s^+ \left\{\pi^- ,\rho^- ,a^-_1
\right\}$ for $B_s$ decays $[W=D^{(*)} D_s^{(*)-}, D\overline D\;\overline K X, J/\psi \overline K^*$ for
$B_d$ decays].

The data samples for the $I(t)$ asymmetries exist already. For instance,
the SLD collaboration determined the lifetime ratio of neutral to charged
$b$-hadrons by an inclusive topological vertex
analysis~\cite{sldtopology}. The polarization of $Z^0$ provides a large forward-backward asymmetry of $b$ production and thus an effective
initial flavor-tag~\cite{polarizedZ0} and it is clear that SLD can study
inclusive asymmetries. Similarly, the LEP experiments are able to study
$I(t)$ by using their $b$-enriched samples and optimal flavor-tagging
algorithms. CDF has several million high $P_T$-leptons, which are highly
enriched in $b$ content. The data sample of detached vertices on the other
hemisphere allows CDF to study $I(t)$. The newly installed vertex detector
at CLEO permits meaningful studies, because the $I(t)$
asymmetry becomes significant only after a few $B_d$ lifetimes, see Eq.
(\ref{incl}) below.

For $\Delta\Gamma =0$, the explicit time dependence is~\cite{bbdcp}  
\begin{equation} 
\label{incl} 
I(t)=a\left[\frac{x}{2} \sin \Delta mt-\sin^2 \frac{\Delta mt}{2}\right]\;, 
\end{equation}
where $x \equiv \Delta m/\Gamma$.
The observable $a$ can thus be extracted from a study of $I(t)$.

For $B_s$ mesons, that extraction offers a significant statistical gain over the conventional method [Eq.~(\ref{wrongsign})].
The factor of $x/2$ enhances $I(t)$ over $a$ by an order of
magnitude, which corresponds to a statistical gain of ${\cal O}
(10^2)$. There is another gain, because all $B_s$ decays are used rather than flavor-specific $B_s$ modes that must be efficiently distinguished from $B_d$ modes.  The distinction involves stringent selection criteria.  The reason is that the wrong-sign asymmetry [Eq.~(\ref{wrongsign})] is time-independent, and the wrong-sign $B_d$ asymmetry is an order of magnitude larger than the $B_s$ one, within the CKM model.  Thus, for instance, the high-p (-$P_T)$ leptons must originate from $B_s$ decays and not from $B_d$ decays. This can be achieved by either studying wrong-sign $B_s$ modes at very short proper times~\cite{jimack}, or by inferring the existence of a $D_s$, or by observing such primary kaons that significantly enrich the $B_s$ content, or by a combination of the above.   In contrast, the unique time-dependence of $I(t)$ provides automatic discrimination.  For the $B_s$ meson at least, the time-dependent inclusive asymmetry may be more effective in extracting the CP violating observable $a$ than the conventional wrong-sign asymmetry.

\begin{figure}
\epsfysize = 3in
\centerline{\vbox{\epsfbox{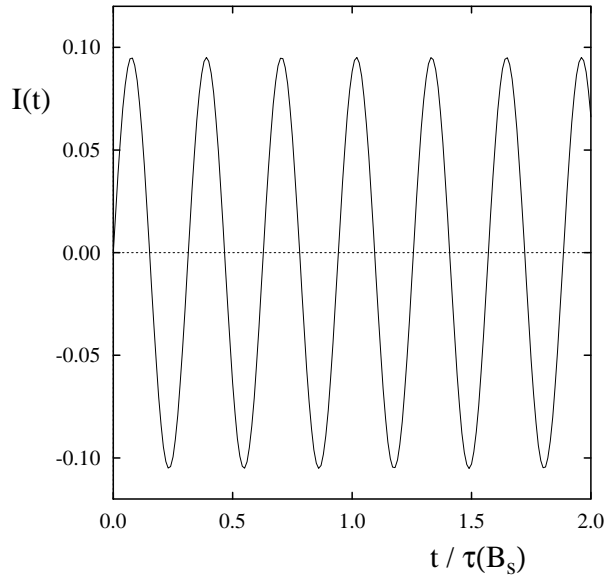}}} \caption{The totally inclusive
CP asymmetry of $B_s(t) \to$ all, with $a= 0.01, \Delta \Gamma = 0$ and $x
= 20$ (see Eqs.~(\protect\ref{incldef}),(\protect\ref{incl})).}
\end{figure}

Figure 1 shows what to expect for the choice $x = 20$ and where New Physics is allowed to enhance
$a=|\Gamma_{12}/M_{12}|\sim 0.01$. The observation of a non-vanishing $I(t)$ proves CP violation
and in addition allows a determination of the $B_s -\overline B_s$ mixing
parameter $\Delta m$ from flavor-nonspecific final states. The traditional
methods for extracting $\Delta m$ require flavor-specific final states and
tagging~\cite{snowmass,xlep}.  We will mention later on additional ways to
extract $\Delta m$ with flavor-nonspecific final states. 

Within the CKM model, the totally inclusive asymmetries are tiny ${\cal
O}(10^{-3})$ for $B_d\;$ and ${\cal O}(10^{-4})$ for
$B_s$~\cite{lusignoli,buchalla}. The ability to select specific quark
transitions enhances the asymmetries by orders of magnitude, at times to
the $\sim (10-20)\%$ level~\cite{bbdcp}. Such selections permit extractions of
CKM phases and to conduct the study in either a time-integrated or
time-dependent fashion.\footnote{For $B_s$ mesons, $\Delta m$ could be
extracted from such more refined studies.} Those analyses should be
pursued whenever feasible.  There exist unitarity constraints, which allow
systematic cross-checks.  Future $B$ detectors will be able to more fully
explore the potential with such semi-inclusive data samples. 

\section{Physics with (untagged) $B_s$ Mesons}

One conventional way to determine the CKM angle $\gamma$ is the
time-dependent study of tagged $\stackrel{(-)}{B_s}(t)\rightarrow D_s^\pm
K^\mp$ processes~\cite{aleksandk}, and in the neglect of penguin
amplitudes $\stackrel{(-)}{B_s}(t)\rightarrow \rho^0 K_S, \omega K_S$
transitions~\cite{dudw,dunietzr,azimov}.  It requires flavor-tagging and
the ability to trace the rapid $\Delta mt$-oscillations.  The requirements
are problematic: 

(a) Flavor-tagging is at present only a few percent efficient at hadron
accelerators~\cite{cdftagging}.\footnote{Though, in principle almost all
$B$-decays could be flavor-tagged~\cite{distinguish}.}

(b) Resolution of $\Delta mt$-oscillations is feasible for $x
\;\raisebox{-.4ex} {\rlap{$\sim$}} \raisebox{.4ex}{$<$} 20$ with present
vertex technology \cite{snowmass}, but LEP experiments reported
\cite{xlep}, 
\begin{equation} 
x \;\raisebox{-.4ex}{\rlap{$\sim$}}
\raisebox{.4ex}{$>$} 15 \;.  
\end{equation}

Though $\Delta mt$-oscillations may be too rapid to be resolved at
present, such large $\Delta m$ may imply a sizable width difference
$\Delta\Gamma$~\cite{bbdbs}.  Non-perturbative effects may further enhance
$\Delta \Gamma$ considerably~\cite{disy}.  Perhaps $\Delta\Gamma$ will be
the first observable $B_s-\overline B_s$ mixing effect~\cite{untagged},
which would circumvent problems (a) and (b). The $\Delta mt$-terms cancel in
the time-evolution of untagged $B_s$~\cite{untagged}, 
\begin{equation}
f(t)\equiv\Gamma (B_s (t) \rightarrow f)+\Gamma (\overline B_s
(t)\rightarrow f)=ae^{-\Gamma_L t}+be^{-\Gamma_H t}\;, 
\end{equation}
which is governed by the two exponentials $e^{-\Gamma_L t}$ and
$e^{-\Gamma_H t}$ alone.  That fact permits many non-orthodox CP violating
studies and extractions of CKM parameters~\cite{untagged}: 

(1) Consider final states with definite CP parity, $f_{CP}$, such as
$\rho^0 K_S, \omega K_S, ....$ If the untagged time-evolution $f_{CP}(t)$
is governed by both exponentials $e^{-\Gamma_L t}$ and $e^{-\Gamma_H t}$,
then CP violation has occured~\cite{untagged}. The measurement of
$f_{CP}(t)$ allows even the extraction of CKM parameters
\cite{untagged,psiphi}.  The physics of the $J/\psi\phi$ final state is
very instructive. The time-evolution of untagged $J/\psi\phi$ could show
CP violating effects \cite{psiphi}. The $\stackrel{(-)}{B_s}\rightarrow
J/\psi\phi$ has CP-even and CP-odd amplitudes, $\stackrel{(-)}{A}_+$ and
$\stackrel{(-)}{A}_-$ respectively. Angular correlations~\cite{dighe}
allow to measure the interference terms between CP-even and CP-odd
amplitudes, which for untagged data samples is proportional
to~\cite{psiphi}, 
\begin{equation} \left(e^{-\Gamma_H t}-e^{-\Gamma_L t}
\right) \theta^2 2\eta \;,\;\;{\rm where}\;\;\;\; \theta \approx 0.22.  
\end{equation} 
The observation of such a
non-vanishing term would prove CP violation and would permit the
extraction of the CKM parameter $\eta$. Note that the observable depends
optimally on the width difference. 

Those interference terms once tagged allow the measurement of $\Delta m$,
even though $J/\psi\phi$ is a flavor-nonspecific final state~\cite{dighe}.
To demonstrate the point most sharply, neglect CP violation and set
$\Delta\Gamma =0$. Then $A_+(t) \sim e^{-im_L t}$ and $A_-(t)\sim e^{-im_H
t}$. The observable $A_+(t) A^*_- (t)\sim e^{i\Delta mt}$ depends on
$\Delta m \equiv m_H - m_L$. Ref.~\cite{azimovd} describes yet another method for measuring $\Delta m$ without flavor-specific final states.

(2) After several $B_s$ lifetimes, the long-lived $B^H_s~\sim
B_s~-\overline B_s$ will be significantly enriched over the short-lived
$B^L_s$. Consider then final states $f$ that can be fed from both $B_s$
and $\overline B_s$, and that are non-CP-eigenstates. CP violation is
proven if the time evolution of untagged $f(t)$ differs from untagged
$\overline f(t)$, 
\begin{equation} 
f(t) \neq \overline f(t) \Rightarrow
{\text CP}\; {\text violation} \;.  \end{equation} 
Furthermore, the CKM
angle $\gamma$ can be extracted from time-dependent studies of $D^\pm_s
K^\mp(t), \stackrel{(-)}{D^0}\phi (t)$~\cite{untagged}.\footnote{The
determination of $\gamma$ from $\stackrel{(-)}{D^0}\phi (t)$ and $D^0_{CP}
\phi (t)$ as presented in Ref.~\cite{untagged} must include the effect of
doubly-Cabibbo suppressed $\stackrel{(-)}{D^0}$
decay-amplitudes~\cite{stone,ads}.} CP violating effects and CKM
extractions can be enhanced by studying $D_s^{(*,**)\pm} K^{*\mp}
(t)$~\cite{fleischerd}. In summary, neither flavor-tagging nor exquisite
tracing of $\Delta mt$-oscillations are necessary, only a large
$\Delta\Gamma$. 

\section{Direct CP Violation and Extracting CKM Angles}

The favorite method (particularly at $\Upsilon (4S)$ factories) for
determining $\gamma$ has been developed by Gronau, London and Wyler
(GLW)~\cite{glw} and requires the measurements of the six rates $B^\pm
\rightarrow D^0 K^\pm ,\overline D^0 K^\pm$ and $D^0_{CP} K^\pm$. Here
$D^0_{CP}$ denotes that the $D^0$ is seen in CP eigenstates with either
CP-even ($K^+K^-,\pi^+ \pi^-,...$) or CP-odd ($K_S\phi ,K_S \pi^0 ,...)$
parity.  The GLW method focuses on the CP violating rate difference of
$B^+\rightarrow D^0_{CP}K^+$ versus $B^-\rightarrow D^0_{CP}
K^-$~\cite{bigisandad0k}, which can reach at best the 10\% level and is
probably significantly smaller. 

In principle, the GLW method is a great idea. However, new CLEO data
indicate that the method is unfeasible, and that the largest CP violating
effect has been overlooked~\cite{stone,ads}. Once the effect has been
incorporated, the CKM angles can be cleanly extracted~\cite{ads}. 

Let us review the original GLW method, point out the problem, and show how
it can be overcome. Consider CP even $D^0_{CP}$, for which
\begin{equation} 
D^0_{CP} =\frac{1}{\sqrt{2}} (D^0 +\overline D^0) \;. 
\end{equation} 
Then 
\begin{equation} \sqrt{2} A(B^-\rightarrow D^0_{CP}
K^-) =A(B^-\rightarrow D^0 K^-) + A(B^-\rightarrow \overline D^0 K^-)\;,
\end{equation} 
and that amplitude triangle is shown in Figure 2. The weak
phase difference of the two interfering amplitudes is $\gamma$. GLW argued
that the magnitudes of each of the sides of the triangle can be measured
(being proportional to the square roots of the respective rates), and thus
claimed that the amplitude triangle can be fully reconstructed. 

\begin{figure}
\vspace*{8.0cm} \includegraphics{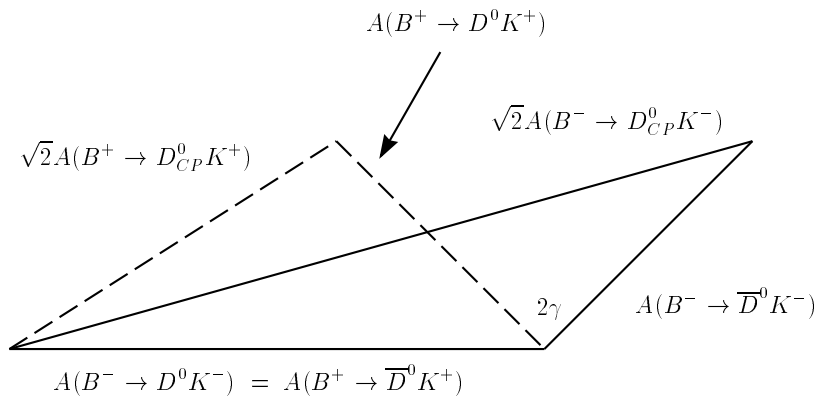} \caption{The traditional GLW method for
extracting the CKM angle $\gamma$.} \end{figure}

Figure 2 has not been drawn to scale. The $B^- \rightarrow \overline D^0
K^-$ amplitude is an order of magnitude smaller than the $B^-\rightarrow
D^0 K^-$ one, which can be seen as follows~ \cite{ads}. The CKM factors
suppress the amplitude ratio by about 1/3. The $\overline D^0 K^-$ is
color-suppressed while $D^0 K^-$ is also color-allowed, yielding another
suppression factor of about 1/4. 

Nothing changes when the CP conjugated final states are considered, except
that the CKM elements have to be complex conjugated.  Apparently, the
CP-conjugated triangle can also be determined, see Figure 2. The
$A(B^+\rightarrow D^0 K^+)$ is rotated by $2\gamma$ with respect to $A(B^-
\rightarrow \overline D^0 K^-),$ and apparently the angle $\gamma$ can be
extracted.  Note that the only CP violation in all these processes occurs
in 
\begin{equation} \Gamma(B^+\rightarrow D^0_{CP}K^+)\neq\Gamma
(B^-\rightarrow D^0_{CP}K^-) \;  
\end{equation} 
while there is no CP
violation in 
\begin{equation} 
\Gamma (B^+\rightarrow \overline D^0 K^+) =
\Gamma (B^-\rightarrow D^0K^-) \;, \;{\text and} 
\end{equation}
\begin{equation} \Gamma (B^+\rightarrow D^0 K^+)=\Gamma (B^-\rightarrow
\overline D^0 K^-)\;.  
\end{equation} 
In principle this argument is
correct, but in practice the largest \underline{direct} CP violating
effects (residing in those processes) will be seen in \cite{stone,ads}
\begin{equation} B^+\rightarrow D^0 K^+ \neq B^-\rightarrow \overline D^0
K^- \;.  
\end{equation} 
The $\overline D^0$ produced in the
$B^-\rightarrow \overline D^0 K^-$ process is seen in its non-leptonic,
Cabibbo-allowed modes $f$, such as $K^+\pi^-, K\pi\pi$.  It was assumed
that the kaon flavor unambiguously informs on the initial charm flavor. 
This assumption overlooked the doubly-Cabibbo-suppressed $D^0\rightarrow
f$ process which leads to the same final state $B^-\rightarrow
D^0[\rightarrow f]K^-$. Further, CLEO has measured~\cite{cleodcs}
\begin{equation} 
\left |\frac{A(D^0 \rightarrow f)}{A(\overline D^0
\rightarrow f)}\right | \sim 0.1\;, 
\end{equation} 
which maximizes the
interference, 
\begin{equation} 
\label{ratioampl} \left |\frac{A(B^-
\rightarrow K^- D^0 [\rightarrow f])}{A(B^- \rightarrow K^- \overline D^0
[\rightarrow f])}\right | \sim 1 \;, 
\end{equation} 
\begin{equation}
A(B^-\rightarrow K^- [f]) = A(B^-\rightarrow K^- D^0 [\rightarrow f]) +
 A(B^- \rightarrow K^- \overline D^0 [\rightarrow f]) \;.  
\end{equation}
The conditions are ideal for striking direct CP violating effects. They
require that the interfering amplitudes be comparable in size (Eq.
(\ref{ratioampl})), that the weak phase difference be large ($\gamma$ in
our case), and that the relative final-state-phase difference be
significant.  It is an experimental fact that large final state phases
occur in many $D$ decays \cite{frabetti}. This enables us to engineer
large CP violating effects by optimally weighting relevant sections of
generalized Dalitz plots. 

The traditional focus on CP eigenmodes of $D^0_{CP}$ automatically
excludes this so potent source of final-state interaction phases. The
orthodox method~\cite{bigisandad0k,glw} accesses only the final-state
phase difference residing in $B^-\rightarrow D^0 K^-$ versus
$B^-\rightarrow \overline D^0 K^-$, which is expected to be significantly
more feeble~\cite{cahns}.  The CKM angle $\gamma$ can be cleanly extracted
once one incorporates the findings of this section~\cite{ads}, because
penguin amplitudes are absent. The extraction of $\gamma$ and the
observation of CP violation is optimized by combining detailed
(experimental) investigations of $D^0$ decays with $B^\pm$ decays to
$\stackrel{(-)}{D^0}$ \cite{ads}. This provides yet another reason for
accurate measurements of $D^0$ decays. Note also that observation of
direct CP violation (as advocated in this section) would rule out
superweak scenarios as the only source for CP violation. 

\section{Conclusion}

CP violation has been observed only in $K^0$ decays and is parameterizable
by a single quantity $\epsilon$. It is one of the necessary ingredients
for baryogenesis \cite{sakharov}, and within the CKM model is related to
the quark-mixing and hierarchy of quark masses.  It is one of the least
understood phenomena in high energy physics and a very important one. Just
as the successful $V-A$ theory of parity violation \cite{vminusa} emerged
from a synthesis of many independent parity violating measurements, so a
more fundamental understanding of CP violation will profit from many
independent observations of CP violation. 

This talk thus emphasized that CP violation should not only be searched in
traditional exclusive $B_d \rightarrow J/\psi K_S ,\pi^+\pi^-$ rate
asymmetries.  Observable CP violating effects could be present in
(semi-)inclusive $B$ decays, and could be searched for with existing data
samples. The time-evolutions of untagged $B_s$ data samples have no rapid
$\Delta mt$-oscillations. Still CP violation could be observed and CKM
parameters extracted as long as $\Delta\Gamma$ is sizable. Many striking
direct CP violating effects in $B$ decays are possible.  The observation
of CP violation and CKM extraction are optimized by detailed studies of
$D$ decays.

\section{Acknowledgements}

This work was supported in part by the   Department of Energy, Contract No.
DE-AC02-76CH03000.



\begin{references}

\bibitem{winsteinw}
B. Winstein and L. Wolfenstein, Rev. Mod. Phys. {\bf 65}, 1113 (1993).

\bibitem{sakharov}
A.D. Sakharov, JETP Lett.~{\bf 5}, 24 (1967).

\bibitem{vminusa}
R.P. Feynman and M. Gell-Mann, Phys. Rev. {\bf 109}, 193 (1958);
E.C.G. Sudarshan and R. Marshak, Phys. Rev. {\bf 109}, 1860 (1958).

\bibitem{buchallak}
G. Buchalla, hep-ph/9612307.

\bibitem{burasf}
For a review see, for instance, A.J. Buras and R. Fleischer, hep-ph/9704376.

\bibitem{sachs}
I. Dunietz and R.G. Sachs, Phys. Rev.~{\bf D 37}, 3186 (1988); (E) ibid.~{\bf D 39}, 3515 (1989).

\bibitem{bbdcp}
M. Beneke, G. Buchalla and I. Dunietz, Phys. Lett. {\bf B393}, 132 (1997).

\bibitem{stodolsky}
L. Stodolsky, hep-ph/9612219.

\bibitem{snowmass} Proceedings of the Workshop on $B$ Physics at Hadron
Accelerators, Snowmass, Co., June 21 - July 2, 1993, edited by P. McBride
and C. Shekhar Mishra. 

\bibitem{xlep}
V. Andreev et al. (The LEP $B$ Oscillations Working Group), "Combined Results on $B^0$ Oscillations: Update for the Summer 1997 Conferences," LEPBOSC 97/2, August 18, 1997.

\bibitem{untagged}
I. Dunietz, Phys. Rev. {\bf D52}, 3048 (1995).

\bibitem{glw}
M. Gronau and D. London, Phys. Lett. {\bf B253}, 483 (1991);
M. Gronau and D. Wyler, Phys. Lett. {\bf B265}, 172 (1991).

\bibitem{ads}
D. Atwood, I. Dunietz and A. Soni, Phys. Rev. Lett. {\bf 78}, 3257 (1997).

\bibitem{stone}
I. Dunietz, Z. Phys. {\bf C56}, 129 (1992);
I. Dunietz, in B Decays, Revised 2nd Edition, edited by S. Stone
(World Scientific, Singapore, 1994), p. 550.

\bibitem{bigisanda}
I.I. Bigi and A.I. Sanda, Nucl. Phys. {\bf B193}, 85 (1981).

\bibitem{wolfenstein}
L. Wolfenstein, Nucl. Phys. {\bf B246}, 45 (1984).

\bibitem{dudw}
D. Du, I. Dunietz and Dan-di Wu, Phys. Rev. {\bf D34}, 3414 (1986).

\bibitem{dunietzr}
I. Dunietz and J.L. Rosner, Phys. Rev. {\bf D34}, 1404 (1986).

\bibitem{dunietzn}
I. Dunietz and T. Nakada, Z. Phys.  {\bf C36}, 503 (1987).

\bibitem{paistreiman}
A. Pais and S.B. Treiman, Phys. Rev. {\bf D12}, 2744 (1975);
 T. Altomari, L. Wolfenstein
and J.D. Bjorken, Phys. Rev. {\bf D 37}, 1860 (1988);
M. Lusignoli, Z. Phys. {\bf C41}, 645 (1989).

\bibitem{hitoshi}
H. Yamamoto, Phys. Lett. {\bf B401}, 91 (1997).

\bibitem{sldtopology}
K. Abe et al. (SLD Collaboration), Phys. Rev. Lett.  {\bf 79}, 590 (1997).

\bibitem{polarizedZ0}
W.B. Atwood, I. Dunietz and P. Grosse-Wiesmann,
Phys. Lett. {\bf B216}, 227 (1989);
W.B. Atwood, I. Dunietz, P. Grosse-Wiesmann, S. Matsuda and
A.I. Sanda, Phys. Lett. {\bf B232}, 533 (1989).

\bibitem{jimack}
M. Jimack, private communication.

\bibitem{lusignoli}
M. Lusignoli, Z. Phys. {\bf C41}, 645 (1989).

\bibitem{buchalla}
G. Buchalla, private communication.

\bibitem{aleksandk}
R. Aleksan, I. Dunietz and B. Kayser, Z. Phys. {\bf C54}, 653 (1992).

\bibitem{azimov}
Ya.I. Azimov, N.G. Uraltsev and V.A. Khoze, JETP Lett. {\bf 43}, 409 (1986).

\bibitem{cdftagging} B. Wicklund, in the proceedings of the b20
conference, June 29 - July 2, 1997, Illinois Institute of Technology,
Chicago, Illinois. 

\bibitem{distinguish}
I. Dunietz, FERMILAB-PUB-94/163-T, hep-ph/9409355.

\bibitem{bbdbs}
M. Beneke, G. Buchalla and I. Dunietz,  Phys. Rev. {\bf D54}, 4419 (1996).

\bibitem{disy} I. Dunietz, J. Incandela, F.D. Snider, and H. Yamamoto,
FERMILAB-PUB-96-421-T (hep-ph/9612421), to be published in Z. Phys. {\bf C}. 

\newpage

\bibitem{psiphi}
R. Fleischer and I. Dunietz, Phys. Rev. {\bf D55}, 259 (1997).

\bibitem{dighe} A.S. Dighe, I. Dunietz, H.J. Lipkin and J.L. Rosner,
Phys. Lett. {\bf B369}, 144 (1996). 

\bibitem{azimovd}
Ya. Azimov and I. Dunietz, Phys. Lett. {\bf B395}, 334 (1997).


\bibitem{fleischerd}
R. Fleischer and I. Dunietz, Phys. Lett. {\bf B387}, 361 (1996).

\bibitem{bigisandad0k}
I.I.Y. Bigi and A.I. Sanda, Phys. Lett. {\bf 211B}, 213 (1988).

\bibitem{cleodcs}
H. Yamamoto, Harvard University report, HUTP-96-A-001, January 1996
[hep-ph/9601218];
D.~Cinabro et al. (CLEO Collab.), Phys. Rev. Lett.
72, 1406 (1994).

\bibitem{frabetti}
See, for instance, P.L. Frabetti (E687 Collaboration), Phys. Lett.
{\bf B331}, 217 (1994);
G. Bonvicini et al. (CLEO Collaboration), contributed paper to the  
28th International
Conference on HEP, Warsaw, Poland, July 1996, PA05-090 [CLEO CONF  
96-21].

\bibitem{cahns}
R.N. Cahn and M. Suzuki, hep-ph/9708208.


\end{references}
\end{document}